\def\be{\begin{equation}}
\def\ee{\end{equation}}
\def\ba{\begin{array}}
\def\ea{\end{array}}

\documentclass[aps,amsmath,amssymb,amsfonts,showpacs,twocolumn,pra]{revtex4}
\usepackage{graphicx}
\usepackage{caption2}
\def\qed{\leavevmode\unskip\penalty9999 \hbox{}\nobreak\hfill
     \quad\hbox{\leavevmode  \hbox to.77778em{%
               \hfil\vrule   \vbox to.675em%
               {\hrule width.6em\vfil\hrule}\vrule\hfil}}
     \par\vskip3pt}

\newtheorem{theorem}{Theorem}
\newtheorem{corollary}{Corollary}

\begin{document}
\title{Improved lower and upper bounds for entanglement of formation}
\author{Xue-Na Zhu$^{1}$}
\author{Shao-Ming Fei$^{2,3}$}

\affiliation{$^1$Department of Mathematics, School of Science, South
China University of Technology, Guangzhou 510640, China\\
$^2$School of Mathematical Sciences, Capital Normal
University, Beijing 100048, China\\
$^3$Max-Planck-Institute for Mathematics in the Sciences, 04103
Leipzig, Germany}

\begin{abstract}

We provide analytical lower and upper bounds for entanglement of
formation for bipartite systems, which give a direct relation
between the bounds of entanglement of formation and concurrence, and
improve the previous results. Detailed examples are presented.

\end{abstract}
\maketitle

Quantum entanglement \cite{R. Horodecki} is of special importance in
quantum-information processing and is responsible for many quantum
tasks such as quantum teleportation \cite{T,T2}, dense coding
\cite{C}, swapping \cite{S,S2}, error correction \cite{E,E2} and
remote state preparation \cite{R,R2}. The entanglement of formation
(EoF) \cite{C. H. Bennett,D. P. DiVincenzo} is a well-defined
important measure of entanglement for bipartite systems.

Let $H_A$ and $H_B$ be $m$- and $n$-dimensional $(m\leq n)$ vector
spaces, respectively. A pure state $|\psi\rangle\in H_A\otimes H_B$
has a Schmidt decomposition
$|\psi\rangle=\sum_{i=1}^m\sqrt{\mu_{i}}|ii\rangle,$  where
$\mu_i\geq0$ and $\sum_{i=1}^m\mu_i=1$. The entanglement of
formation is given by the entropy of the reduced density matrix
$\rho_A = Tr_B(|\psi\rangle\langle\psi|)$,
\begin{equation}\label{S}
E(|\psi\rangle)=S(\rho_A)=-\sum_{i=1}^{m}\mu_i\log\mu_i\equiv H(\vec \mu),
\end{equation}
where log stands for the natural logarithm throughout the paper,
$\mu_i~(i=1, 2, ..., m)$ are the non-zero eigenvalues of $\rho_A$
and $\vec \mu$ is the Schmidt vector $(\mu_1,\mu_2, . . . ,\mu_m)$.
For a bipartite
 mixed state $\rho$, the entanglement of formation is
given by the minimum average marginal entropy of the ensemble decompositions of $\rho$,
\begin{equation}\label{EE}
E(\rho)=\min_{\{p_i,|\psi_i\rangle\}}\sum_ip_iE(|\psi_i\rangle),
\end{equation}
for all possible ensemble realizations $\rho=\sum_ip_i|\psi_i\rangle \langle \psi_i|$,
where $p_i\geq0$ and $\sum_ip_i=1$.

Another significant measure of quantum entanglement is the concurrence. The concurrence of
a pure bipartite state $|\psi\rangle$ is given by
\begin{equation}\label{CON}
C(|\psi\rangle)=\sqrt{2[1-Tr(\rho^2_A)]}=\sqrt{2(1-\sum_{i=1}^m\mu^2_i)}.
\end{equation}
It is extended to mixed states by the convex roof construction
\begin{equation}\label{CONC}
C(\rho)=\min_{\{p_i,|\psi_i\rangle\}} \sum_i p_i C(|\psi_i\rangle),
\end{equation}
for all possible ensemble realizations $\rho=\sum_i p_i |\psi_i\rangle \langle \psi_i|$.

Considerable effort has been made to estimate
the entanglement of formation and concurrence for bipartite quantum
states, and their lower and upper bounds
via analytical and numerical approaches.
For the two-qubit case, EoF is a monotonically increasing function of
the concurrence, and an analytical formula of concurrence
has been derived \cite{W. K. Wootters}.
For the general high-dimensional case, due to the extremizations involved in
the computation, only a few analytic formulas have been obtained
for isotropic states \cite{B. M. Terhal} and Werner states \cite{K. G. H}
for EoF, and for some special symmetric
states \cite{Terhal-Voll2000,Terhal-Voll20002,Terhal-Voll20003} for concurrence.

Instead of analytic formulas, some progress has been
made toward the lower and upper bounds of EoF and concurrence for any
$m\otimes n$ $(m\leq n)$  mixed quantum state $\rho$.
In \cite{K.Chen1,K.Chen2,F. Mintert, J. I. deVicente,LiX}, explicit analytical lower and upper
bounds of concurrence have been presented.
In Ref.\cite{K.Chen1}, a simple analytical lower bound of EoF
has been derived. Recently new results related to the
bounds of EoF have been further derived in \cite{LiMing,CH}.
In this article, we give new lower and upper bounds of EoF based on the concurrence.
Detailed examples are presented, showing that our bounds improve the bounds
in \cite{LiMing,CH}.

In Ref.\cite{LiMing}, the authors defined $X(\lambda)$  and
$Y(\lambda)$ to derive measurable lower and upper bounds of EoF. We
give an improved definition of $X(\lambda)$ and $Y(\lambda)$ in this
paper. For a given pure state
$|\psi\rangle=\sum_{i=1}^m\sqrt{\mu_{i}}|ii\rangle$, the concurrence
of $|\psi\rangle$ is given by $c=\sqrt{2(1-\sum_{i}^{m}\mu_i^2)}$.
However, the pure states with the same value of concurrence $c$ are not unique.
Namely, different sets of the Schmidt coefficients $\{\mu_i\}$ may give rise to
the same concurrence $c$.
The entanglement of formation $H(\vec\mu)$ defined in Eq.(\ref{S}) for a pure state
depends on the Schmidt coefficients $\{\mu_i\}$.
We define the maximum and minimum of $H(\vec\mu)$
to be $X(c)$ and $Y(c)$ for fixed $c$,
\begin{equation}\label{x}
X(c)=\max\left\{H(\vec\mu)\left\vert\sqrt{2(1-\sum_{i=1}^m\mu^2_i)}\equiv
c\right.\right\},
\end{equation}
and
\begin{equation}\label{y}
Y(c)=\min\left\{H(\vec\mu)\left\vert\sqrt{2(1-\sum_{i=1}^m\mu^2_i)}\equiv
c\right.\right\},
\end{equation}
respectively, where the maximum and minimum are taken
over all possible Schmidt coefficient distributions $\{\mu_i\}$
such that the value of concurrence $c=\sqrt{2(1-\sum_{i}^{m}\mu_i^2)}$ is fixed.

Let $\varepsilon(c)$ be the largest monotonically increasing convex function that is bounded above
by $Y(c)$, and $\eta(c)$ be the smallest monotonically increasing concave function that is
bounded below by $X(c)$.

\begin{theorem}\label{th1}
For any $m\otimes n$ $(m\leq n)$ quantum state $\rho$, the
entanglement of formation $E(\rho)$ satisfies
\begin{equation}\label{th}
\varepsilon(C(\rho))\leq E(\rho)\leq\eta(C(\rho)).
\end{equation}
\end{theorem}

$Proof.$ We assume that $\rho=\sum_{i}p_i|\psi_i\rangle\langle\psi_i|$ is the optimal decomposition of $E(\rho)$.
Therefore
\begin{equation}
\ba{rcl}
E(\rho)&=&\sum_{i}p_iE(|\psi_i\rangle)=\sum_ip_iH(\vec \mu_i)\\[3mm]
&\geq&\sum_{i}p_i\varepsilon(c_i)\geq\varepsilon(\sum_{i}p_ic_i)\\[3mm]
&\geq&\varepsilon(C(\rho)).
\ea
\end{equation}
We have used the definition of $\varepsilon(c)$ to obtain the first
inequality. The second inequality is due to the convex property of
$\varepsilon(c)$, and the last one is derived from the definition of
concurrence.
On the other hand, as $\rho=\sum_{i}p_i|\psi_i\rangle\langle\psi_i|$ is the optimal
decomposition of $C(\rho)$, we have
\begin{equation}
\ba{rcl}
E(\rho)&\leq&\sum_{i}p_iE(|\psi_i\rangle)=\sum_ip_iH(\vec \mu_i)\\[3mm]
&\leq&\sum_{i}p_i\eta(c_i)\leq\eta(\sum_{i}p_ic_i)\\[3mm]
&=&\eta(C(\rho)),
\ea
\end{equation}
where we have used the definition of $E(\rho)$ to
obtain the first inequality. The second and the third inequalities are due to the definition of $\eta(c)$.
\qed

Our analytical bounds (\ref{th}) give an explicit relations between
the EoF and the concurrence. In fact, if we denote $\underline{c}$ and $\overline{c}$ the analytical
lower and upper bounds of concurrence, respectively,
according to Theorem \ref{th1}, we have the following corollary:

\begin{corollary}
For any $m\otimes n (m\leq n)$ quantum state $\rho$, the
entanglement of formation $E(\rho)$ satisfies
\begin{equation}\label{co}
\varepsilon(\underline{c})\leq E(\rho)\leq\eta(\overline{c}).
\end{equation}
\end{corollary}

Here $\underline{c}$ and $\overline{c}$ could be any known analytical
lower and upper bounds of concurrence. For example, from the known bounds of concurrence
in \cite{K.Chen1,LiMing,CH}, one may choose
\begin{widetext}
 $\underline{c}=\max\left\{0,\sqrt{\frac{2}{m(m-1)}}(||\rho^{T_A}||-1 ),\sqrt{\frac{2}{m(m-1)}}
 \left[R(\rho)-1\right],\sqrt{2[Tr(\rho^2)-Tr(\rho^2_A)]},\sqrt{2[Tr(\rho^2)-Tr(\rho^2_B)]}\right\}$
\end{widetext}
and $\overline{c}=\min\left\{\sqrt{2[1-Tr(\rho^2_A)]},\sqrt{2[1-Tr(\rho^2_B)]}\right\}$,
where $\rho ^{T_{A}}$ stands for the partial transpose with respect to the
subsystem $A$, $R(\rho)$ is the realigned matrix of $\rho$, $||\cdot ||$ stands for the trace norm, and
$\rho_A$ and $\rho_B$ are the reduced density matrices with respect to the subsystems $A$ and $B$ respectively.

The maximal admissible $H(\vec\mu)$ and the minimal admissible
$H(\vec\mu)$ in Eqs.(\ref{x}) and (\ref{y}) for a given $c$ can be
estimated following the approach in Ref.\cite{LiMing}. Let $n_1$ be the
number of entries such that $\mu_i=\alpha$ and let $n_2$ be the number
of entries such that $\mu_i=\beta$. The maximal admissible
$H(\vec\mu)$ and the minimal admissible $H(\vec\mu)$ for a given $c$
become, for fixed $ n_1,n_2,n_1 +n_2 \leq m$ , one of maximizing or
minimizing the function
\begin{equation}\label{F}
F_{n_1,n_2}=n_1h(\alpha_{n_1n_2})+n_2h(\beta_{n_1n_2}),
\end{equation}
where $h(x)=-x\log x$,
$$
\alpha_{n_1n_2}=\frac{n_1+\sqrt{n^2_1-n_1(n_1+n_2)[1-n_2(1-\frac{c^2}{2})]}}{n_1(n_1+n_2)},
$$
and $\beta_{n_1n_2}=({1-n_1\alpha_{n_1n_2}})/{n_2}$.

When $m=3$, to find the expressions of upper and lower
bounds in Eqs.(\ref{th}) and (\ref{co}) is to obtain the maximization and minimization
over the three functions $F_{11}(c)$, $F_{12}(c)$ and $F_{21}(c)$. From Eq.(\ref{F}), for $m=3$, we have
\begin{equation}\label{x1}
X(c)=\left\{
\begin{aligned}
&F_{11}(c),~~~ 0< c\leq1,\\
&F_{12}(c),~~~ 1< c\leq\frac{2}{\sqrt{3}},
\end{aligned}
\right.
\end{equation}
and
\begin{equation}\label{y1}
Y(c)=\left\{
\begin{aligned}
&F_{11}(c),~~~ 0<c\leq1,\\
&F_{21}(c),~~~ 1< c\leq\frac{2}{\sqrt{3}}.
\end{aligned}
\right.
\end{equation}

To determine $\varepsilon(c)$ and $\eta(c)$, we  study the
concavity and convexity of functions  $F_{11}(c), F_{12}(c)$ and
$F_{21}(c)$.  Since $F_{11}^{''}\geq0$ and $F_{21}^{''}\leq0$, where
$F_{ij}^{''}$ is the second derivative of $F_{ij}$, we have that
$\varepsilon(c)$ is the curve consisting of $F_{11}$ for $c\in(0,1]$
and the line connecting the points $[1,F_{21}(1)]$ and
$[\frac{2}{\sqrt{3}},F_{21}(\frac{2}{\sqrt{3}})]$ for
$c\in(1,\frac{2}{\sqrt{3}}]$, that is,
\begin{equation}\label{var}
\varepsilon(c)=\left\{
\begin{aligned}
&F_{11}(c),~~~ 0<c\leq1,\\
&\frac{\sqrt{3}\log3/2}{2-\sqrt{3}}(c-1)+\log2,~~~ 1< c\leq\frac{2}{\sqrt{3}}.
\end{aligned}
\right.
\end{equation}
Similarly, since $F_{11}^{''}\geq0$ and $F_{12}^{''}\geq0$, we have
that $\eta(c)$ is the curve connecting the points $[0,0]$ and
$[\frac{2}{\sqrt{3}},F_{12}(\frac{2}{\sqrt{3}})]$ for $c\in(0,1]$,
and the line connecting the points $[1,F_{12}(1)]$ and
$[\frac{2}{\sqrt{3}},F_{12}(\frac{2}{\sqrt{3}})]$ for
$c\in(1,\frac{2}{\sqrt{3}}]$, that is,
\begin{widetext}
\begin{equation}\label{eta}
\eta(c)=\left\{
\begin{aligned}
&\log2 (c),~~~ 0<c\leq1,\\
&\frac{2\log 3/2+\log6-3\log3}{3(\sqrt{3}-2)}(\sqrt{3}\,c-2)+\log3,~~~ 1< c\leq\frac{2}{\sqrt{3}}.
\end{aligned}
\right.
\end{equation}
\end{widetext}
See Fig. 1.
\begin{figure}[htpb]
\renewcommand{\captionlabeldelim}{.}
\renewcommand{\figurename}{FIG.}
\centering
\includegraphics[width=7.5cm]{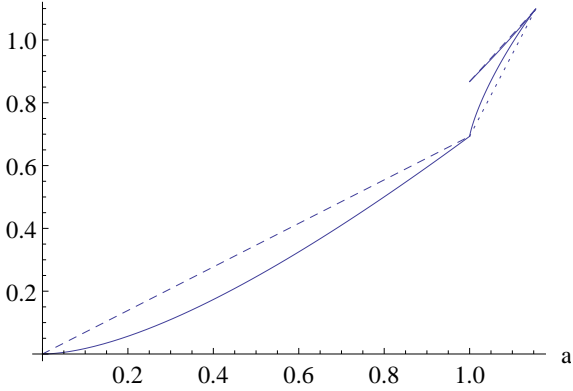}
\caption{{\small Upper and lower bounds of $E(\rho)$ (dished lines and dotted lines ), and $F_{11},F_{12} \text{ and } F_{21}$ (solid lines).}}
\label{2}
\end{figure}

Similarly, for any $m$, we can get the expressions of $X(c)$ and $Y(c)$,
\begin{equation}\label{X}
X(c)=\left\{
\begin{aligned}
&F_{11}(c),~~~ 0< c\leq1,\\
&F_{12}(c),~~~ 1< c\leq\frac{2}{\sqrt{3}},\\
&....\\
&F_{1(m-1)}(c),~~~ \sqrt{\frac{2(m-2)}{m-1}}< c\leq\sqrt{\frac{2(m-1)}{m}},
\end{aligned}
\right.
\end{equation}
and
\begin{equation}\label{Y}
Y(c)=\left\{
\begin{aligned}
&F_{11}(c),~~~ 0< c\leq1,\\
&F_{21}(c),~~~ 1< c\leq\frac{2}{\sqrt{3}},\\
&...\\
&F_{(m-1)1}(c),~~~ \sqrt{\frac{2(m-2)}{m-1}}< c\leq\sqrt{\frac{2(m-1)}{m}}.
\end{aligned}
\right.
\end{equation}

The representations
of $\varepsilon(c)$ and $\eta(c)$ can be also similarly calculated analytically  in accordance with the following principles $(2\leq t\leq m-1)$:

If $F^{''}_{1t}(c)\geq0,
~c\in(\sqrt{\frac{2(t-1)}{t}},\sqrt{\frac{2t}{t+1}}]$, then $\eta(c)=[F_{1t}(\sqrt{\frac{2t}{t+1}})-F_{1t}(\sqrt{\frac{2(t-1)}{t}})]
(c-\sqrt{\frac{2t}{t+1}})/[\sqrt{\frac{2t}{t+1}}-\sqrt{\frac{2(t-1)}{t}}]
+F_{1t}(\sqrt{\frac{2t}{t+1}})$;
 If $F^{''}_{1t}(c)\leq0,
~c\in(\sqrt{\frac{2(t-1)}{t}},\sqrt{\frac{2t}{t+1}}]$, then $\eta(c)=F_{1t}(c)$;
 If $F^{''}_{t1}(c)\geq0,
~c\in(\sqrt{\frac{2(t-1)}{t}},\sqrt{\frac{2t}{t+1}}]$, then $\varepsilon(c)=F_{t1}(c)$;
 If $F^{''}_{t1}(c)\leq0,
~c\in(\sqrt{\frac{2(t-1)}{t}},\sqrt{\frac{2t}{t+1}}]$, then $\varepsilon(c)=[F_{t1}(\sqrt{\frac{2t}{t+1}})-F_{t1}(\sqrt{\frac{2(t-1)}{t}})]
(c-\sqrt{\frac{2t}{t+1}})/[\sqrt{\frac{2t}{t+1}}-\sqrt{\frac{2(t-1)}{t}}]+F_{t1}(\sqrt{\frac{2t}{t+1}})$.


The bounds given in Theorem 1 and Corollary 1 can be used to
improve the bounds of EoF presented in \cite{LiMing} and \cite{CH}.
In fact, the lower bound obtained in Ref.\cite{CH} is better than the lower bound from Ref.\cite{LiMing},
while the upper bounds are the same. Our bounds are obtained from the improved bounding functions
(\ref{x}) and (\ref{y}). They are directly given by the concurrence. From the concurrence, or the lower and upper
bounds of the concurrence of a given mixed state, one can get analytical lower and upper bounds of
EoF of the state.
To see the tightness of inequalities (\ref{th}) and (\ref{co}), let us consider the following examples.

{\it Example. 1} Let us consider the well-known Werner states, which are a class of
mixed states for $d\otimes d$ systems that
are invariant under the transformations $U\otimes U$, for
any unitary transformation $U$ \cite{K. G. H,RT}. The density matrix of the Werner states can be expressed as
\begin{equation}
\rho_{f}=\frac{1}{d^3-d}[(d-f)I+(df-1)\mathcal{F}],
\end{equation}
where $\mathcal{F}$ is the flip operator defined
by $\mathcal{F}(\phi\otimes \psi)=\psi\otimes \phi$.
Consider the case $d=3$.
We have $(\rho_f)_A=(\rho_f)_B=\frac{1}{3}\left(|0\rangle\langle0|+|1\rangle\langle1|+|2\rangle\langle2|\right)$ and
$1-Tr[(\rho_f)^2_A]=\frac{2}{3}$. By Refs.\cite{LiMing,CH},
the upper bound of EoF is given by $E(\rho_f)\leq1.099$.
From Eq.(\ref{eta}), we get the upper bound of $\rho_f$,
\begin{equation}
E(\rho_{f})\leq-f\log2,~~~ -1\leq f<0,
\end{equation}
It is obvious that $-f\log2<1.099$ for $-1\leq f<0$. Hence the upper bound (\ref{eta}) is better than the upper bound in Refs.\cite{LiMing,CH}.

{\it Example. 2} Consider the $3\otimes3$ mixed state
$\rho=\frac{x}{9}I+(1-x)|\psi\rangle \langle \psi|$, where the column vector
$|\psi\rangle=(a,0,0,0,\frac{1}{\sqrt{3}},0,0,0,\frac{1}{\sqrt{3}})^t/\sqrt{a^2+2/3}$
with $a\in[0,1]$, $t$ stands for vector transposition. For this state we have
$Tr(\rho^2)-Tr(\rho^2_A)=Tr(\rho^2)-Tr(\rho^2_B)=2[9-26 x + 9a^4
(-2+ x)x+13x^2+6a^2(9-22x+11x^2)]/9 (2 + 3 a^2)^2$ and
$1-Tr(\rho^2_A)=1-Tr(\rho^2_B)=[6+4x-18a^4(-2+x)x-2x^2+12a^2(3-2x+x^2)]/{3(2+3a^2)^2}$.

For fixed $x=0.1$, one has
\begin{equation}
\ba{rcl}
\underline{c}&=&\sqrt{2[Tr(\rho^2)-Tr(\rho^2_A)]}\\[3mm]
&=&\displaystyle\frac{2\sqrt{6.53+41.46a^2-1.71a^4}}{3(2+3a^2)},
\ea
\end{equation}
\begin{equation}
\ba{rcl}
\overline{c}&=&\sqrt{2[1-Tr(\rho^2_A)]}\\[3mm]
&=&\displaystyle\frac{1}{2+3a^2}\sqrt{\frac{2(6.38+33.72a^2+3.42a^4)}{3}}.
\ea
\end{equation}
Substituting $\underline{c}$ and $\overline{c}$ into Eqs.(\ref{var}) and (\ref{eta}),
we have the upper and lower bounds for $E(\rho)$, see Fig. 2.
\begin{figure}[htpb]
\renewcommand{\captionlabeldelim}{.}
\renewcommand{\figurename}{FIG.}
\centering
\includegraphics[width=7.5cm]{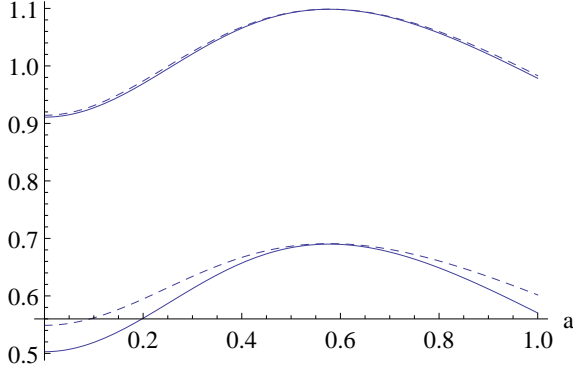}
\caption{{\small Upper and lower bounds of $E(\rho)$ when x=0.1. Dished lines are given by Eqs.(\ref{var}) and (\ref{eta}}), solid lines are given by Ref. \cite{LiMing}.}
\label{2}
\end{figure}

Similarly for $x=0.001$, one has
$\underline{c}$ and $\overline{c}$, and the upper and lower bounds for $E(\rho)$, see Fig. 3.
Form Figs. 2 and 3, we see that the lower bound of EoF (\ref{var}) is better than
the one from \cite{LiMing}.
\begin{figure}[htpb]
\renewcommand{\captionlabeldelim}{.}
\renewcommand{\figurename}{FIG.}
\centering
\includegraphics[width=7.5cm]{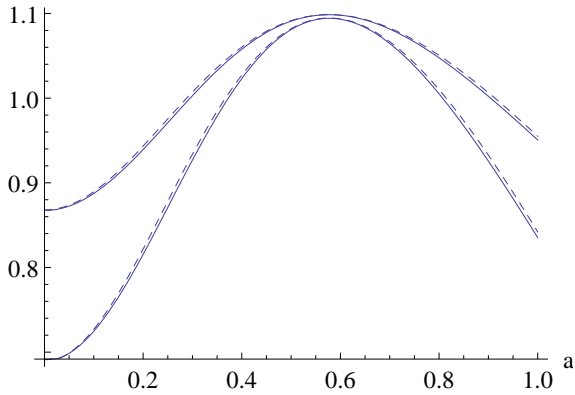}
\caption{{\small Upper and lower bounds of $E(\rho)$ when x=0.001. Dished lines are given by Eqs.(\ref{var}) and (\ref{eta}}), solid lines are given by Ref. \cite{LiMing}.}
\label{2}
\end{figure}

In fact, we can make the bounds better by choosing suitable $\underline{c}$ and $\overline{c}$.
For instance, to find a better lower bound, we may choose $\underline{c}=\sqrt{\frac{1}{3}}(||\rho^{T_A}||-1)$.
Then for $x=0.1$ and $a\in[0.5,0.66]$, one has
\begin{equation}\label{p1}
||\rho^{T_A}||-1=\frac{2[5+6.9a^2-0.9a^4+9.353a(2+3a^2)]}{3(2+3a^2)^2}.
 \end{equation}
When $ x=0.001$ and $a\in[0.57,0.59]$, one has
\begin{equation}\label{p2}
||\rho^{T_A}||-1=\frac{2[5.99+8.978a^2-0.009a^4+10.382a(2+3a^2)]}{3(2+3a^2)^2}.
 \end{equation}
Substitute Eqs.(\ref{p1}) and (\ref{p2}) into (\ref{var}), respectively,
we get another lower bound of EoF for the state in example 2, see Figs. 4 and 5.
\begin{figure}[htpb]
\renewcommand{\captionlabeldelim}{.}
\renewcommand{\figurename}{FIG.}
\centering
\includegraphics[width=7.5cm]{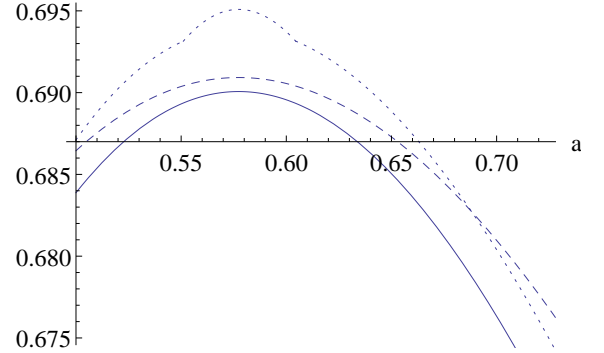}
\caption{{\small Lower bounds of $E(\rho)$ when x=0.1. Dashed line is obtained by Ref.\cite{CH}, dotted line is obtained by Eq.(\ref{var}) and  solid line is obtained by Ref.\cite{LiMing}.}}
\end{figure}
\begin{figure}[htpb]
\renewcommand{\captionlabeldelim}{.}
\renewcommand{\figurename}{FIG.}
\centering
\includegraphics[width=7.5cm]{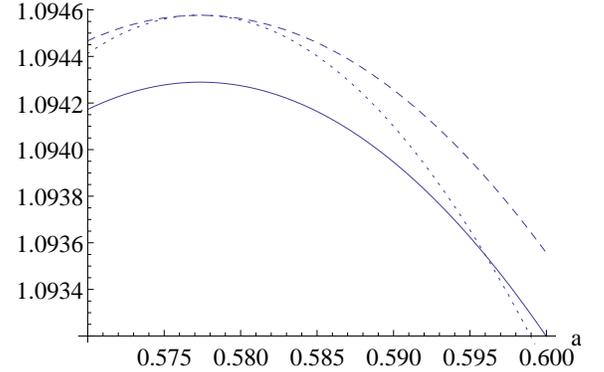}
\caption{{\small Lower bounds of $E(\rho)$ when x=0.001. Dashed line is obtained by Ref.\cite{CH}, dotted line is obtained by Eq.(\ref{var}) and  solid line is obtained by Ref.\cite{LiMing}.}}
\end{figure}

From Figs. 4 and 5, it is obvious that
$\varepsilon(\sqrt{\frac{1}{3}}(||\rho^{T_A}||-1))>\varepsilon_{1}>\varepsilon_0$
when $x=0.1$ and $a\in[0.5,0.66]$, and
$\varepsilon_{1}=\varepsilon(\sqrt{2[Tr(\rho^2)-Tr(\rho^2_A)]})>\varepsilon(\sqrt{\frac{1}{3}}(||\rho^{T_A}||-1))>\varepsilon_0$
when $x=0.001$ and $a\in[0.57,0.59]$, where $\varepsilon_0$ and
$\varepsilon_1$ are obtained by Ref.\cite{LiMing} and Ref.\cite{CH},
respectively . The lower bounds are improved in the particular
interval of $a$.

The density matrix in example 2 is close to being separable (pure) when $x$ is close to $1$ (0). To show better
the advantage of our results, we now take $x=0.6$. We have then
$Tr(\rho^2)-Tr(\rho_A^2)=Tr(\rho^2)-Tr(\rho_B^2)<0$, and hence the
lower bounds from Refs.\cite{LiMing,CH} are
$\varepsilon_0=\varepsilon_1=0$. However, by choosing
\begin{widetext}
\begin{equation}
\underline{c}=\frac{||\rho^{T_A}||-1}{\sqrt{3}}
=\frac{2(6+a^2(9-21x)+6\sqrt{3}a(2+3a^2)(x-1)-10x-9a^4x)}{3\sqrt{3}(2+3a^2)^2},
\end{equation}
\end{widetext}
we have $\underline{c}>0$ when $a\in(0.205,1)$. From Corollary
$1$ we get that $E(\rho)\geq\varepsilon(\underline{c})>0$, see Fig.6 for the lower bound $\varepsilon(\underline{c})$,
\begin{figure}[htpb]
\renewcommand{\captionlabeldelim}{.}
\renewcommand{\figurename}{FIG.}
\centering
\includegraphics[width=7.5cm]{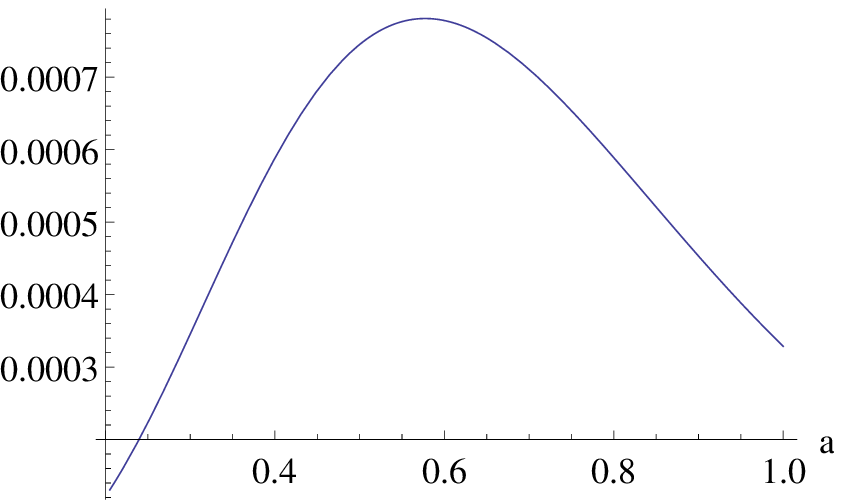}
\caption{{\small Lower bounds of $E(\rho)$ when x=0.6.}}
\end{figure}

In summary, we have presented analytic
lower and upper bounds of EoF for arbitrary bipartite mixed states.
The bounds can be used to improve the previous results on bounds of EoF.
Although the EoF is a monotonically increasing function of the concurrence only
in the two-qubit case, it turns out that for higher dimensional cases, the bounds of EoF have
a tight relation to the concurrence or the bounds of concurrence.

\bigskip
\noindent{\bf Acknowledgments}\, This work is supported by the NSFC
11275131 and PHR201007107.

\end{document}